\documentclass[referee, usenatbib, useAMS]{biom}
\usepackage{graphicx}
\usepackage{bm, amsmath}
\begin{document}

\title[Melding Wildlife Surveys]
{Melding Wildlife Surveys to Improve Conservation Inference}

\author{\\ Justin J. Van Ee$^{1,*}$\email{vanee002@colostate.edu},
Christian A. Hagen$^{2}$, David C. Pavlacky Jr.$^{3}$, Kent  A. Fricke$^{4}$, \\ Matthew D. Koslovsky$^1$,  and Mevin B. Hooten$^5$ \\
$^{1}$Department of Statistics, Colorado State University, Fort Collins, CO \\
$^{2}$Department of Fisheries, Wildlife, and Conservation Sciences, Oregon State University, Corvallis, OR \\
$^{3}$Bird Conservancy of the Rockies, Brighton, CO \\
$^{4}$Kansas Department of Wildlife and Parks, Emporia, KS \\
$^{5}$Department of Statistics and Data Sciences, The University of Texas at Austin, Austin, TX}

 \begin{abstract}
     Integrated models are a popular tool for analyzing species of conservation concern. Species of conservation concern are often monitored by multiple entities that generate several datasets. Individually, these datasets may be insufficient for guiding management due to low spatio-temporal resolution, biased sampling, or large observational uncertainty. Integrated models provide an approach for assimilating multiple datasets in a coherent framework that can compensate for these deficiencies. While conventional integrated models have been used to assimilate count data with surveys of survival, fecundity, and harvest, they can also assimilate ecological surveys that have differing spatio-temporal regions and observational uncertainties. Motivated by independent aerial and ground surveys of lesser prairie-chicken, we developed an integrated modeling approach that assimilates density estimates derived from surveys with distinct sources of observational error into a joint framework that provides shared inference on spatio-temporal trends. We model these data using a Bayesian Markov melding approach and apply several data augmentation strategies for efficient sampling. In a simulation study, we show that our integrated model improved predictive performance relative to models that analyzed the surveys independently. We use the integrated model to facilitate prediction of lesser prairie-chicken density at unsampled regions and perform a sensitivity analysis to quantify the inferential cost associated with reduced survey effort. 
 \end{abstract}
 
 \begin{keywords}
conservation biology; data augmentation; integrated modeling; lesser prairie-chicken; Markov melding.
\end{keywords}
 
\maketitle

\section{Introduction}
 
Integrated models that allow for the unified analysis of multiple datasets have been described as integrated analysis \citep{Maunder2013}, integrated distribution models \citep{Isaac2020}, shared parameter models \citep{Rizopoulos2008}, joint models \citep{Wulfsohn1997}, Markov combination \citep{Dawid1993}, Bayesian melding \citep{Fuentes2005}, data assimilation \citep{Ghil1991}, data reconciliation \citep{Hanks2011}, and data fusion \citep{Kedem2017} and have applications in econometrics, biostatistics, conservation biology, atmospheric sciences, and oceanography. The joint likelihood of integrated models conditions multiple datasets on link parameters in a way that can often improve predictive performance and parameter precision \citep{Schaub2011}. One difficulty with specifying the joint likelihood, however, is choosing the link parameters such that they may be related across datasets but differ in interpretation. 

Markov combination \citep{Dawid1993} facilitates joint inference on a link parameter expressed in several submodels but is not applicable when the prior marginal distributions of the link parameter differ across submodels. \cite{Goudie2019} introduced Markov melding for combining related submodels that have differing interpretations for the link parameter. In this setting, the joint model is constructed through marginal replacement, where the prior marginal distributions for the link parameter across submodels are replaced with a common pooled prior distribution. Markov melding facilitates joint inference on a link parameter in one submodel that can be expressed as non-invertible functions of other submodel parameters. For example, suppose we have submodels for learning about adult and juvenile survival, but we are interested in learning about aggregate survival, which is a weighted average of the two. Markov melding uses marginal replacement to form a melded posterior distribution for the link parameter that accounts for its implied prior and likelihood in each submodel. Recently, \cite{Manderson2022a} proposed chained Markov melding, an extension that facilitates joint inference for a sequence of submodels connected by multiple link parameters. 
 
The earliest applications of integrated modeling frameworks in the context of wildlife management arose in fisheries science \citep{Fournier1982}, but wide adoption of the framework in the broader fields of conservation biology and ecology began in the early 2000s \citep{Maunder2013, Zipkin2018}. In particular, integrated population models (IPMs), which are an application of integrated models, have been used to understand population dynamics for species of conservation concern \citep{Schaub2011, Zipkin2018}. In an IPM, population counts are analyzed in conjunction with surveys of survival, fecundity, and harvest by conditioning all datasets on a shared latent process that describes population dynamics \citep{Schaub2011, Zipkin2018, Schaub2021}. For example, \cite{Broms2010} specified an IPM for greater sage-grouse (\textit{Centrocercus urophasianus}) that leveraged count, telemetry, and harvest data to understand drivers of abundance. By accounting for uncertainty in multiple datasets, IPMs can provide novel insights into population dynamics that help inform conservation \citep{Schaub2021}. 

Despite the success of IPMs, few other integrated modeling approaches have been proposed in conservation biology. One persistent challenge is the lack of spatial and temporal conformity across datasets. Additional methodological challenges include differences in the quantity or observational uncertainty of the data sources, and sampling bias in one or more datasets \citep{Isaac2020, Simmonds2020, Zipkin2021}. Such challenges are encountered when developing integrated models for species of conservation concern (SCC) because of their elusiveness, restricted range, or small population size \citep{Lomba2010}. 

We developed an integrated model that facilitates joint inference of aerial and ground surveys of lesser prairie-chicken (\textit{Tympanuchus pallidicinctus}; hereafter LEPC), an SCC that has experienced range and population declines since the 1980s \citep{Hagen2004, Hagen2017, USFWSAssessment2021}. Joint modeling of these data is challenging because LEPC are simultaneously monitored by several entities who operate independently in different regions. As a result, the surveys vary in their spatial and temporal resolutions, sample size, and observational uncertainties. Additionally for some surveys, LEPC were preferentially sampled in regions presumed to have high abundances which may bias inference \citep{Diggle2010}.

We facilitated shared inference of multiple LEPC surveys by chained Markov melding \citep{Manderson2022a} density estimates derived from submodels describing the observation processes of the aerial and ground surveys into a joint response model. Melding refines the submodel density estimates to those that agree with the spatio-temporal patterns observed in both surveys. By joining the submodels through derived quantities, we addressed the differences in the spatial and temporal scales of the surveys. Accommodating these differences in scales with a traditional integrated model is difficult because density is a non-invertible function of submodel parameters. Our modeling approach attenuated the impacts of potential sampling biases and accounted for the distinct sources of observational error so that all data sources can be assimilated to improve predictive performance. Lastly, the Markov melding approach improved computation by enabling submodel specific data augmentation techniques and avoiding high-dimensional parameter updates by fitting the integrated model in stages. 

The paper is organized as follows. In Section \ref{LEPCbackground}, we provide a brief history of LEPC conservation and discuss current needs for informing management. Section \ref{datacollection} details the sampling protocols of the aerial (\ref{dataaerial}) and ground (\ref{dataground}) surveys. Section \ref{Models} describes submodels accounting for the observation process of each survey and a joint response model for linking inference across surveys. In Section \ref{modelingfitting}, we describe the Markov melding techniques used to facilitate posterior inference for our integrated model. Section \ref{results} includes the results and a simulation study and sensitivity analysis to access the predictive performance of our integrated model. Section \ref{discussion} concludes with a discussion of our findings. 

\section{Lesser prairie-chicken Conservation}\label{LEPCbackground}
 
The LEPC is a member of the family Phasianidae and is indigenous to the southern Great Plains of the United States. Like other species in its family, the LEPC has experienced range and population declines since the 1980s primarily due to habitat loss, degradation, and fragmentation \citep{Hagen2004, Hagen2017, USFWSAssessment2021}, but curtailment of natural fires, overgrazing, and climate change have also contributed \citep{Haukos2016}. 

We studied spatio-temporal patterns in LEPC abundance across the state of Kansas because an estimated 70\% of the total LEPC population resides in the state \citep{Van2013}. Our modeling approach, however, can accommodate data sources from the other states in the LEPC range. In Kansas, LEPC inhabit Sand Sagebrush Prairie (SSPR), Mixed Grass Prairie (MGPR), Shortgrass Prairie/Conservation Reserve Program Mosaic (SGPR) ecoregions, which cover the southwest, southeast, and northern regions of western Kansas respectively. 

Recently, the United States Fish and Wildlife Service listed the LEPC for federal protections under the Endangered Species Act \citep{USFWS2022LEPC}. The Northern Distinct Population Segment, which encompasses the SSPR, MGPR, and SGPR ecoregions, is categorized as threatened. Improved estimation of spatio-temporal population change, especially range-wide, would help inform conservation practices for the species \citep{Van2013}.
 
Population monitoring of LEPC relies on spring counts of individuals on leks \citep{McDonald2014}. A lek is an aggregation of males defending a small territory and communally calling and performing displays to attract and mate with females \citep{Haukos2016}. Leks are generally located in sparse vegetation on hilltops and ridgelines and commonly include more than 10 individuals which makes detection by audio and visual cues of the otherwise cryptic individuals easier \citep{Haukos2016}.

Historically, LEPC populations have been monitored using counts of individuals at leks from ground surveys conducted by state wildlife agencies. Lack of spatial randomness in the ground surveys, however, makes inferring species-habitat associations difficult and density estimates imprecise and potentially biased \citep{Diggle2010}. Since 2012, several entities have collectively supported annual range-wide aerial surveys of LEPCs. The aerial surveys follow a spatially random sampling design and have thereby improved range-wide density estimates \citep{McDonald2014, Nasman2021}. Two drawbacks of the aerial surveys is that they encounter fewer individuals per unit of area searched and have higher operating costs. These limitations have led managers to consider integrated models that could leverage ground survey data and reduce reliance on aerial surveys. 

Over the last two decades, there have been numerous studies related to LEPC conservation but few have assimilated multiple data sources due to the methodological challenges described by \cite{Zipkin2021}. \cite{Ross2018} developed an IPM for assimilating count, survival, and fecundity data that suggested observed declines in LEPC abundance following droughts \citep{Ross2016} were driven by higher juvenile and chick mortality. The findings of \cite{Ross2018} prompted managers to consider habitat improvements that focus on increasing and maintaining grasslands that can buffer the population against the harmful effects of severe drought. By melding available data sources, we improve spatio-temporal density estimates and facilitate prediction at unsampled regions to identify vulnerable populations and prioritize landscapes for conservation action. Our approach can also quantify the inferential cost and reduced predictive performance associated with less frequent aerial surveys. In what follows, we describe the aerial and ground survey protocols.

\section{Survey Protocols}\label{datacollection}
 
\subsection{Aerial}\label{dataaerial}
 
\begin{figure}
\includegraphics[scale=0.31]{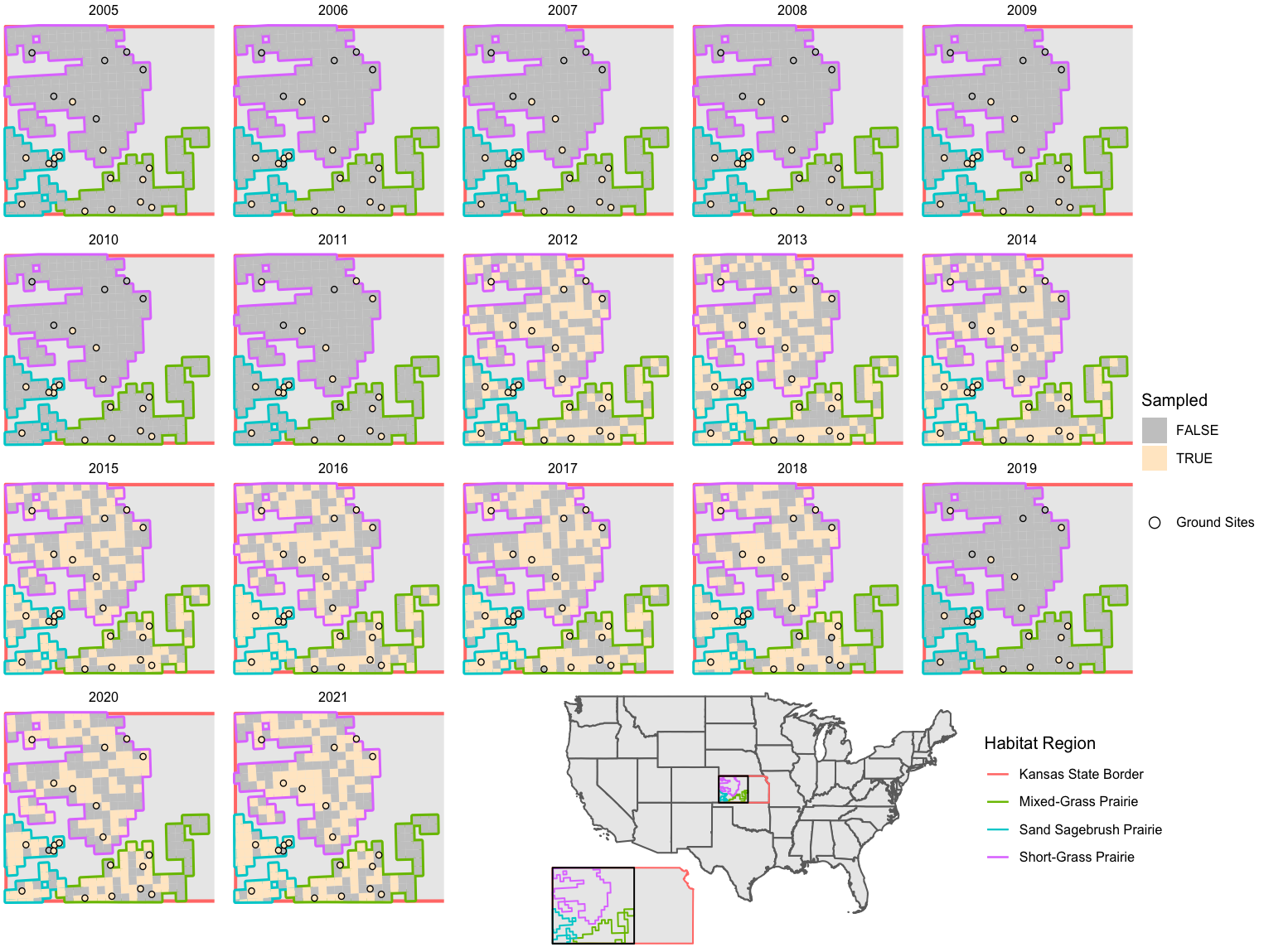} 
    \caption{Map of Kansas LEPC aerial survey blocks and Kansas Department of Wildlife and Parks ground monitoring sites. Golden fill indicates the block/site was sampled during that year. The region encompassed by all three ecoregions in the map is Kansas estimated occupied range.}
  \label{fig:samplesites}
\end{figure}

The Kansas estimated occupied range (EOR) for LEPC was partitioned into $n^{A}=299$, $15\times 15=225$ km$^2$, survey blocks \citep{McDonald2014}. A spatially random subset of blocks were selected for sampling, and the subset selected differed by year (Figure \ref{fig:samplesites}). No blocks were surveyed in 2019. 

Two north-south oriented, 15-km transects were surveyed by helicopter in blocks selected for sampling. Selected transects were surveyed once during the LEPC breeding season (March 15-May 15) and within 0.5 hours prior to and 2 hours after sunrise to maximize detection of individuals present at leks. The helicopter was operated by one pilot and three observers. As the pilot flew at a speed of 60 km per hour and altitude of 25 meters above ground, observers attempted to visually locate prairie-chicken. When one or more prairie-chicken were located, the pilot navigated to the location and recorded the geographic coordinate and number of individuals observed. For an in-depth description of the aerial survey protocol and design, see \cite{Nasman2021} and \cite{Van2013}.
 
\subsection{Ground}\label{dataground}
 
Kansas Department of Wildlife and Parks (KDWP) preferentially located 21 ground survey routes for monitoring LEPC in representative, high quality LEPC habitat across Kansas EOR. Each route was approximately 16 km long and the ground survey attempted to census all leks within 1.6 kilometers of the road for a region of approximately 51.2 km$^2$. Routes were surveyed (March 20-April 20) and within 0.5 hours prior to and 1.5 hours after sunrise. 
 
All routes were surveyed at least twice per year in two parts. First, the listening portion of the route was conducted; leks were audibly detected and their locations approximated, but not confirmed. On the same morning, the surveyor navigated to each lek detected, prompted the individuals to take flight (flushed), and recorded the count of individuals and location. Surveyors also revisited sites at which leks were previously recorded because LEPC are known to return to historical lek sites \citep{Haukos2016}. The ground survey is a census of the leks in the survey area but it is not a census of the population because some individuals may not be present at their lek at the time it was flushed. 

\section{Methods}\label{Models}

In northwestern Kansas, the LEPC EOR overlaps with the range of its sister species the greater prairie-chicken (\textit{Tympanuchus cupido}; hereafter GEPC). Species verification was sometimes infeasible for the aerial and ground surveys and observations of GEPC are included in both datasets. We proposed distance sampling (Section \ref{Aerialmodel}) and N-mixture (Section \ref{N-mixture}) submodels that analyzed counts of prairie-chicken (LEPC and GEPC). We then derived the block-level densities of LEPC in northwestern Kansas by multiplying the combined LEPC and GEPC density estimates by known LEPC to GEPC ratios (Section \ref{STTM}). The spatio-temporal submodel assimilates the LEPC density estimates derived from the other two submodels in a joint response that induced the integrated model. The integrated model accounted for the uncertainty in both datasets, the underlying ecological processes, and the parameters.

 \subsection{Aerial Distance Sampling Submodel}\label{Aerialmodel}
 
 We developed a distance sampling model to describe the observational uncertainty associated with aerial surveys of prairie-chickens. We let $v_{itl}$ represent the number of observers who detected group $l=1,\dots,L^{A}_{it}$ in sampling region $i=1,\dots,n^{A}$ during year $t=1,\dots,T^{A}$. Assuming all observers had equal skill in detecting prairie-chicken groups and observers detected the groups independently, a model for $v_{itl}$ is
 \begin{align}
     v_{itl} \sim \text{Binomial}(B_{itl}, \rho_{itl}),
 \end{align}
 where $B_{itl}$ is the total number of observers for which group $l$ was visible and $\rho_{itl}$ is the observer detection probability for group $l$, assumed to be identical for all observers. The visibility of group $l$ to each observer depended on their distance from the transect, $d_{itl}$, and side of the transect, $\epsilon_{itl}$ ($\epsilon_{itl}=1$ indicates group on left side). Groups more than 7 meters left of the transect were visible to both the front and rear left-hand side observers; groups within 7 meters of the transect were only visible to the front left-hand side observer; and groups more than 7 meters right of the transects were only visible to the right-hand observer. Hence, $B_{itl}=2$ for $\epsilon_{itl}=1$ and $d_{itl}>7$, but $B_{itl}=1$ otherwise. Detected prairie-chicken groups were announced only after they were out of view for all observers to ensure independent detections. 
 
We modeled the detection probability of group $l$, $\rho_{itl}$, as a function of the group's distance from the transect at detection, $d_{itl}$, count of individuals at detection, $N^{A}_{itl}$, and ecoregion, such that $\text{logit}(\bm{\rho})=\left(\bm{X}_{\rho}, \bm{N}^{A}, \bm{d}\right)\bm{\beta}_{\rho}$, where $\bm{X}_{\rho}$ is a binary matrix with unique intercepts for each ecoregion, and $(\cdot)$ denotes a column-wise bind of the listed matrices. The regression model provides additional flexibility for estimating the detectability of prairie-chicken groups, and the entries of $\bm{\beta}_{\rho}$ are identifiable under the double observer design \citep{Borchers2006}. We treat detections of the two left-hand observers as fully independent but alternative approaches that allow for dependence in detectability as a result of unmeasured covariates and animal movement have been proposed \citep{Buckland2010, Borchers2022}. Under our modeling framework, we assume that heterogeneity in prairie-chicken group detectability is well characterized by distance from the transect and size of the group. We also assumed groups are stationary, but note that there were a small number of transiting individuals.

Some groups for which $v_{itl}=0$ were not in the dataset because they went undetected. To account for these missed individuals, we employed a parameter expanded data augmentation (PX-DA) approach \citep{Royle2009}. Specifically, we augmented the dataset with many undetected groups and let $z_{itl}\in\{0,1\}$ indicate whether group $l$ belonged to the sample population of groups in region $i$. If a group was detected (i.e., $v_{itl}>0$), then it must be part of the sample population in region $i$ (i.e., $z_{itl}=1$). 

For undetected groups, $z_{itl}$, $N^{A}_{itl}$, $d_{itl}$, and $\epsilon_{itl}$ were all unknown and hence estimated. To denote the observed and unobserved components of partially latent parameters, we use the superscripts $o$ and $u$, respectively. Heuristically, we conceptualize the model as proposing groups of prairie-chicken that the aerial survey may have missed; we proposed a group of prairie-chicken with count $N^{A, u}_{itl}$, distance from the transect $d^{u}_{itl}$, and on side $\epsilon^{u}_{itl}$ of the transect, and then used the observations from our detected groups (i.e., $\bm{N}^{A, o}$, $\bm{d}^{o}$, $\bm{\epsilon}^{o}$) to determine if group $l$ could have been part of our sample population (i.e., $z^{u}_{itl}=1$) but went undetected (i.e., $v_{itl}=0$). We chose the prior distributions for $d_{itl}$ and $\epsilon_{itl}$ to induce a uniform distribution of groups within the survey region. See Web Appendix A for a full description of prior distributions.  \cite{Royle2009} referred to the total number of both observed and unobserved groups as the super population, and the size of the super population, $M$, must be specified \textit{a priori}. Web Appendix B discusses recommendations for choosing $M$. We calculate the total number of groups in the sample population of region $i$ during year $t$ as the derived quantity $L^{A}_{it}=\sum_{l=1}^{M}z_{itl}$. Note that in this data augmentation framework $L^{A}_{it}$ includes the detected groups as well as groups that may have existed in the survey region but went undetected. 

The aerial survey was conducted during the breeding season to maximize detection of leks, but smaller, non-lekking groups as well as individual prairie-chicken were also detected. We accounted for the occurrence of lek and non-lek observations in the observed prairie-chicken counts using a zero-truncated Poisson (ZTP) mixture model
    \begin{align}
 & N^{A}_{itl} \sim  \begin{cases} 
  \text{ZTP}(\lambda_{it}), & \text{ for $\omega_{itl}=1$} \\  
  \text{ZTP}(\lambda_0), & \text{ for $\omega_{itl}=0$} \\
                    \end{cases}, \label{Poissonmixturemodel1} \\ 
 & \omega_{itl} \sim \text{Bernoulli}(p_\omega), \label{Poissonmixturemodel2}
 \end{align}
where $\omega_{itl}$ is the indicator of whether group $l$ is a lek, $\lambda_{it}$ is the mean number of individuals per lek in region $i$ during year $t$, and $\lambda_0$ is the homogeneous mean number of individuals for non-lek observations. Both distributions in the Poisson mixture (\ref{Poissonmixturemodel1})-(\ref{Poissonmixturemodel2}) are zero-truncated because if a group exists, it must have $\ge1$ individuals. 
 
We treated $\omega_{itl}$ as a latent variable because it was often infeasible to determine the lek status of a prairie-chicken group from the air. For monitoring purposes, KDWP defines a lek as 3 or more individuals on a display site \citep{Jennison2011}. In our case, the latent lek indicators $\omega_{itl}$ accommodated the bimodality of the count data and carried fewer assumptions regarding the composition of a lek. 
 
Mean lek size varies temporally and with environmental factors \citep{Hagen2009, Hagen2017}. We specified a heterogeneous mean lek size across sites $i$ and years $t$, $\lambda_{it}$, which we modeled with covariates (i.e., $\log(\bm{\lambda})=\bm{X}_{\lambda}\bm{\beta}_{\lambda}$).
The design matrix $\bm{X}_{\lambda}$ includes unique intercepts for each ecoregion and additional continuous covariates. The covariates capture heterogeniety in mean lek size related to landcover, habitat patch size, anthropogenic disturbance, and climatic stochasticity. See Web Appendix C for a description of all covariates, and how they were collected. 
 
We specified a binomial model to account for variability in the number of prairie-chicken groups such that
 \begin{align}
    & L^{A}_{it} \sim \text{Binomial}(M, \psi_{it}), 
 \end{align}
where $\psi_{it}$ is the probability that a group belonged to the sample population of region $i$ during year $t$. The parameter $\psi_{it}$ controls the number of prairie-chicken groups within a region, with greater $\psi_{it}$ implying more groups. Heterogeneity in prairie-chicken use of habitat within the EOR has also been documented \citep{Hagen2016}, motivating the logit model, $\text{logit}(\bm{\psi}) = \bm{X}_{\psi}\bm{\beta}_{\psi}$.
We chose the same suite of covariates for explaining heterogeneity in the number of groups as those used for explaining lek size $\left(\text{i.e., } \bm{X}_{\psi}=\bm{X}_{\lambda}\right)$.
 
We specified diffuse exchangeable Gaussian priors for the regression coefficients $\bm{\beta}_{\rho}$, $\bm{\beta}_{\lambda}$, and $\bm{\beta}_{\psi}$. We used a vague $\text{Uniform}(0,1)$ prior for the proportion of prairie-chicken groups that are leks, $p_\omega$, and an informative $\text{Gamma}(1.78, 0.675)$ prior for the mean number of individuals for non-lek observations $\lambda_0$.  A full description of the priors is in Web Appendix A.
 
\subsection{N-mixture Submodel}\label{N-mixture}
 
We developed a submodel for describing observational uncertainty in KDWP prairie-chicken ground surveys. We let $F_{itlj}$ denote the ground count of male prairie-chicken on occasion $j$ at lek site $l$ in sampling region $i$ during year $t$. To account for variability in the counts induced by imperfect male lek attendance, we adopted a N-mixture model \citep{Royle2004},
\begin{align}
    F_{itlj} \sim \text{Binomial}(N^{G}_{itl}, p) & \text{ for $j=1,\dots,J_{itl}$},
\end{align}
where $p$ represents the homogeneous probability that a male belonging to lek site $l$ was present at the lek when it is surveyed. We assumed a Poisson model for the latent lek abundances, $N^{G}_{itl}\sim\text{Poisson}(\exp{(\bm{w}_{it}'\bm{\eta})})$, where $\bm{w}_{it}$ is the same set of covariates used in the aerial model but with unique measurements because the aerial and ground sample regions differed. Note that zero abundances, $N_{itl}=0$, were possible because surveyors revisited historical lek sites that may not have been visited by any individuals in year $t$. It follows that $\exp{(\bm{w}_{it}'\bm{\eta})}$ is the expected number of individuals per lek site rather than the expected number of individuals per active lek, and the regression coefficient $\bm{\eta}$ dictates the relationship between the expected number of individuals at a lek site and the covariates associated with that lek site. We specified a diffuse exchangeable Gaussian prior for $\bm{\eta}$ and a vague $\text{Uniform}(0,1)$ prior for the male lek attendance probability $p$ (see Web Appendix A for more details of the prior specification).
 
\subsection{Integrated Model}\label{STTM}
 
We induced an integrated model for the aerial and ground surveys by specifying a spatio-temporal submodel that couples the survey specific density estimates in a joint response. While density is not a parameter in either the ADSM or N-mixture submodel, each submodel includes density as a derived quantity. For the ADSM, samples of block-level LEPC density in the aerial lattice are obtained by 
\begin{align}\label{densityderivationaerial}
y_{it}^{A}=f(\bm{N}^{A, o}_{it}, \bm{z}^{o}_{it}, \bm{N}^{A, u}_{it},\bm{z}^{u}_{it})=\sum_{l=1}^{M}N^{A}_{itl}z_{itl}\kappa_{i}/S^{A},
\end{align}
where $S^{A}$ is the prespecified area of the sampling region (Web Appendix B) and $\kappa_{i}$ is the ratio of LEPC to GEPC in sampling region $i$ \citep{Nasman2022}. Ratios vary from $0.001$-$1$ for blocks in the SGPR but equal 1 for all blocks in the MGPR and SSPR. Likewise, for the N-mixture submodel, 
\begin{align}\label{densityderivationground}
y_{it}^{G}=g(\bm{N}^{G}_{it})=\sum_{l=1}^{L^{G}_{it}}2N^{G}_{itl}\kappa_{i}/S^{G}_i,
\end{align}
where $S_i^{G}$ is area of survey route $i$, $L^{G}_{it}$ is the number of lek sites at site $i$ in year $t$, and the $2$ assumes equal sex ratios in the LEPC population \citep{Campbell1972}. Equation (\ref{densityderivationground}) also assumes no females were present at the time the lek site was flushed which is a common assumption but could lead to inflated estimates of $y_{it}^{G}$. Both $y_{it}^{A}$ and $y_{it}^{G}$ are unobserved because they are functions of, at least partially, unobserved submodel parameters.
 
Given the annual density estimates for the $n^A=299$ aerial blocks arranged in a lattice as well as the $n^G=21$ ground survey routes (Figure \ref{fig:samplesites}), we proposed a joint response model for annual density at the $n^A+n^G=320$ sampling regions. Omitting the superscripts $A$ and $G$, we let $y_{it}$ represent the density of LEPC in sampling region $i$ during year $t$. Because some sampling regions can have a LEPC density of exactly zero, we considered the following tobit model \citep{Amemiya1985}:
 \begin{align}
     &  y_{it} = \begin{cases} 
  \zeta_{it}, & \text{ for $\zeta_{it}>0$} \\
  0, & \text{ for $\zeta_{it}\le0$} \\
                    \end{cases}, \\  
    &  \bm{\zeta}_{t} \sim \mathcal{N}(\bm{\xi}_{t}, \sigma_d^2\bm{R}_d(\phi)).
 \end{align}
 
Tobit models are often used in the context of censoring where the true state of interest, $\bm{\zeta}_{t}$, is only observable in a certain range. Our density data were not censored explicitly, but the tobit model accounted for the mixture of discrete and continuous components in the response and promoted conjugacy of the latent states $\zeta_{it}$ and $\xi_{it}$. Both $\zeta_{it}$ and $\xi_{it}$ may be viewed as the latent density of LEPC in a region with negative values indicating the relative probability that the density is zero. To account for spatial structure, we assumed an exponentially decaying correlation matrix $\bm{R}_d(\phi)$, where the entry in the $i$th row and $j$th column is defined as $r_d(i,j, \phi) = \exp(-d_{ij}/\phi)$, $d_{ij}$ is the Euclidean distance between sampling regions $i$ and $j$ in meters, and $\phi$ is the spatial range parameter. 
 
We accounted for temporal dependence by specifying autoregressive random effects in (9)
 \begin{align}
&  \bm{\xi}_{t} \sim \mathcal{N}(\left(\bm{\xi}_{t-1},\bm{W}_{t-1}\right)\bm{\alpha}, \bm{\Sigma}_{\tau}),
 \end{align}
where $\bm{W}_{t-1}$ is a matrix of covariates measured across all sampling regions in year $t-1$, and we modeled the initial state as $\bm{\xi}_{0} = \bm{X}_0\bm{\gamma}$. Many environmental factors known to be associated with LEPC density were constant over the $T=17$ years considered in our analysis, and so the set of covariates used in $\bm{W}_t$ is reduced from those in $\bm{X}_0$ (see Web Appendix C). In addition to the landcover and climatic covariates used in the ADSM and N-mixture submodel, we also included a binary covariate that indicated whether a survey block or ground site was north of Interstate 70. LEPC to GEPC ratios decrease sharply north of Interstate 70 \citep{Nasman2022}, and the binary covarite was helpful for explaining spatial heterogeneity in LEPC density that was difficult to characterize with the other covariates. We considered a block diagonal structure for $\bm{\Sigma}_{\tau}$ with distinct covariance matrices $\sigma_{\tau}^{2,A}\bm{R}^{A}_{\tau}$ and $\sigma_{\tau}^{2,G}\bm{R}^{G}_{\tau}$ for the aerial and ground survey regions, respectively. We let $\bm{R}^{A}_{\tau}=\left(\text{diag}(\bm{A}\bm{1})-\rho\bm{A}\right)^{-1}$, where $\bm{A}$ is the adjacency matrix from the aerial survey lattice, which has entries $a(i,j)=1$ if blocks $i$ and $j$ are neighboring and $a(i,j)=0$ otherwise, and $\text{diag}(\bm{A}\bm{1})$ denotes the diagonal matrix of the row sums of $\bm{A}$. We specified $\rho\rightarrow1$ to induce an intrinsic conditional autoregressive covariance matrix that allows for dependence among regions organized in a lattice \citep{Ver2018}. For the ground sites, we designated a simple diagonal structure $\bm{R}^{G}_{\tau}=\bm{I}$. We used diffuse exchangeable Gaussian priors for the regression coefficients $\bm{\gamma}$ and $\bm{\alpha}$, a discrete uniform prior for $\phi$, and vague inverse-gamma priors for the variance parameters $\sigma^2_{d}$, $\sigma^{2,A}_{\tau}$, and $\sigma^{2,G}_{\tau}$.
 
\begin{figure}
 \includegraphics[scale=0.36]{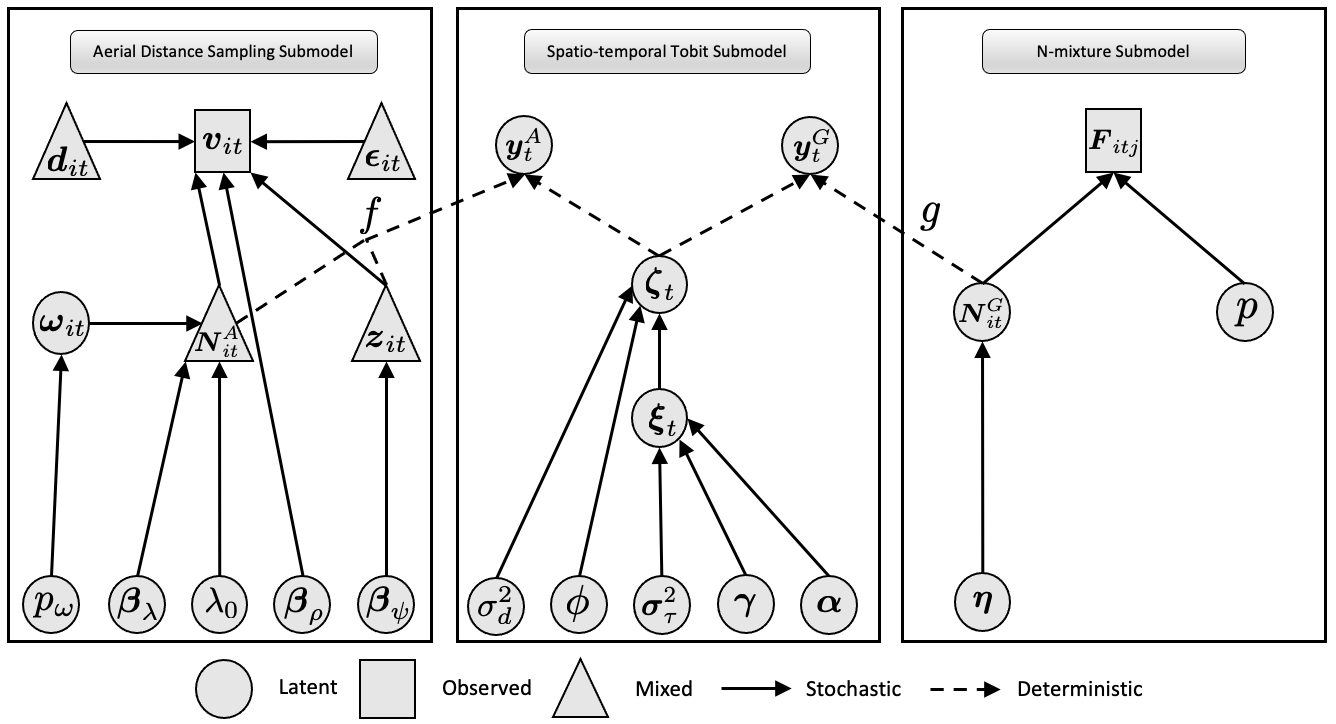} 
\caption{Directed acyclic graph of integrated model. Note that $\bm{\sigma}^2_{\tau}=(\sigma^{2,A}_{\tau}, \sigma^{2,G}_{\tau})'$.}
\label{fig:DAG}
\end{figure}

The joint posterior distribution associated with our full integrated model is 
\small
\begin{align}
& [\bm{\beta}_{\lambda}, \bm{\beta}_{\rho}, \bm{\beta}_{\psi}, \bm{\gamma}, \bm{\alpha}, \bm{\eta}, \bm{\zeta}, \bm{\xi}, \bm{\omega}, p_\omega, p, \sigma^{2,A}_{\tau}, \sigma^{2,G}_{\tau}, \sigma^2_{d}, \phi, \bm{N}^{A,u}, \bm{N}^{G}, \bm{z}^{u}, \bm{d}^{u}, \bm{\epsilon}^{u}  | \bm{N}^{A,o}, \bm{z}^{o}, \bm{d}^{o}, \bm{\epsilon}^{o}, \bm{v}, \bm{F}] \label{posterior}  \\ \propto
& \prod_{t=1}^{T^{G}}{\prod_{i=1}^{n^A}{\prod_{l=1}^{M}{\Bigg( [v_{itl}|N^{A}_{itl}, z_{itl}, d_{itl}, \epsilon_{itl}, \bm{\beta}_{\rho}][N^{A}_{itl}|\bm{\beta}_{\lambda},\lambda_{0},\omega_{itl}][\omega_{itl}|p_\omega][z_{itl}|\bm{\beta}_{\psi}] \Bigg) }}}[\bm{\beta}_{\lambda}][\bm{\beta}_{\rho}][\bm{\beta}_{\psi}][\bm{d}^{u}][\bm{\epsilon}^{u}][\lambda_0][p_\omega] \label{posterioraerial}  \\
 \times & \prod_{t=1}^{T^{G}}\prod_{i=1}^{n^{G}}\Bigg(\prod_{l=1}^{L_{it}^{G}}\Big(\prod_{j=1}^{J_{itl}}[F_{itlj}|N^{G}_{itl},p]\Big)[N^{G}_{itl}|\bm{\eta}]\Bigg)[\bm{\eta}][p] \label{posteriorN-mixture} \\
\times & \prod_{t=1}^{T^{G}}{\Bigg([\bm{N}_t^{G},\bm{N}^{A}_t,\bm{z}_t|\bm{\xi}_t,\sigma^2_{d},\phi][\bm{\xi}_t|\bm{\xi}_{t-1},\bm{\alpha},\sigma^{2,A}_{\tau}, \sigma^{2,G}_{\tau}]\Bigg)}[\bm{\gamma}][\bm{\alpha}][\sigma^2_{d}][\sigma^{2,A}_{\tau}] [\sigma^{2,G}_{\tau}][\phi], \label{posteriorSTTM}
\end{align}
\normalsize
where we use the bracket notation to denote probability distributions \citep{Gelfand1990}. The joint distributions of the ADSM, N-mixture submodel, and spatio-temporal tobit submodel (STTM) are given by (\ref{posterioraerial}), (\ref{posteriorN-mixture}), and (\ref{posteriorSTTM}), respectively. The three submodels induced the integrated model through the link parameters $y_{it}^{A}=f(\bm{N}^{A, o}_{it}, \bm{z}^{o}_{it}, \bm{N}^{A, u}_{it},\bm{z}^{u}_{it})$ and $y_{it}^{G}=g(\bm{N}^{G}_{it})$. A directed acyclic graph of our integrated model is shown in Figure \ref{fig:DAG}, and a full model statement with priors is provided in Web Appendix A. 

\section{Posterior Inference}\label{modelingfitting}

The crux of fitting our integrated model was that the link parameters $\bm{y}^{A}$ and $\bm{y}^{G}$ are non-invertible functions of the submodel parameters $\bm{z}$, $\bm{N}^{A}$, and $\bm{N}^{G}$. We adopted a chained Markov melding approach \citep{Manderson2022a} that facilitated joint inference for $\bm{y}^{A}$ and $\bm{y}^{G}$ accounting for the data, prior information, and assumptions in all three submodels. We derive the joint melded distribution for $\bm{y}=({\bm{y}^A}',{\bm{y}^G}')'$ as follows \citep{Manderson2022a}:
\begin{align}
    [\bm{y},\cdot]_{\text{meld}} & = [\bm{y}]_{\text{pool}}[\cdot|\bm{y}]_{\text{ADSM}}[\cdot|\bm{y}]_{\text{STTM}}[\cdot|\bm{y}]_{\text{N-mix}}, \label{marginalreplacement} \\
    & = [\bm{y}]_{\text{pool}}\frac{[\cdot,\bm{y}^A]_{\text{ADSM}}}{[\bm{y}^A]_{\text{ADSM}}}\frac{[\cdot,\bm{y}]_{\text{STTM}}}{[\bm{y}]_{\text{STTM}}}\frac{[\cdot,\bm{y}^G]_{\text{N-mix}}}{[\bm{y}^G]_{\text{N-mix}}}, \label{simplify} 
\end{align}
where ``$\cdot$'' is a placeholder for all parameters other than $\bm{y}$ in the joint and conditional distributions, $[\bm{y}]_{\text{pool}}$ is the pooled prior marginal distribution, and $[\cdot, \bm{y}]_{\mathcal{M}}$ and $[\bm{y}]_{\mathcal{M}}$ denote the joint and prior marginal distribution of $\bm{y}$ in submodel $\mathcal{M}$, respectively. In the first equality (\ref{marginalreplacement}), we perform marginal replacement to establish a common prior marginal distribution for $\bm{y}$ across all submodels \citep{Goudie2019}. \cite{Goudie2019} proved that marginal replacement minimizes the Kullback–Leibler divergence between the melded distribution and original joint distribution under the constraint that the updated joint distribution admits $[\bm{y}]_{\text{pool}}$ as a marginal. Therefore, we can view (\ref{marginalreplacement}) as the minimally modified joint distribution with marginal $[\bm{y}]_{\text{pool}}$. Note that neither of the conditional distributions $[\cdot|\bm{y}]_{\text{ADSM}}$ or $[\cdot|\bm{y}]_{\text{N-mix}}$ in (\ref{marginalreplacement})  have an analytical closed form because both $f$ and $g$ are non-invertible (\ref{densityderivationaerial}-\ref{densityderivationground}) . We therefore rewrite the joint melded distribution as a product of the submodel joint distributions over the prior marginals for posterior inference (\ref{simplify}).

Another difficulty in posterior inference is that all three submodel marginals in (\ref{simplify}) are analytically intractable. \cite{Goudie2019} recommended approximating the submodel marginals with kernel density estimators, but this approach can lead to numerical instabilities in implementation \citep{Manderson2022b}. We obviated approximating the submodels marginal distribution by 
constructing $[\bm{y}]_{\text{pool}}$ using chained product of experts (PoE) pooling \citep{Manderson2022a},
\begin{align}
    [\bm{y}]_{\text{pool}}=\frac{1}{K}[\bm{y}^{A}]_{\text{ADSM}}[\bm{y}]_{\text{STTM}}[\bm{y}^{G}]_{\text{N-mix}}\text{,  for } K=\int[\bm{y}^{A}]_{\text{ADSM}}[\bm{y}]_{\text{STTM}}[\bm{y}^{G}]_{\text{N-mix}}d\bm{y}.
\end{align}
Under PoE pooling, the melded posterior for $\bm{y}$ is proportional to a product of the submodel joint distributions, which simplifies implementation. One caution regarding PoE is that the pooled prior is often unintuitive and may not be a good summary of the submodel marginals \citep{Goudie2019}. We simulated draws from $[\bm{y}^{A}]_{\text{ADSM}}$, $[\bm{y}]_{\text{STTM}}$, and $[\bm{y}^{G}]_{\text{N-mix}}$ using standard (forward) Monte Carlo methods and found that the implied prior marginals were vague because the specified priors for submodel parameters $\bm{\beta}_{\lambda}$, $\bm{\alpha}$, $\bm{\eta}$, etc., were also vague. Because of the limited impact of prior information and pooling function on posterior inference for $\bm{y}$, we used PoE pooling for computational convenience, but see \citep{Goudie2019} for a suite of other pooling options.

Targeting $[\bm{y},\cdot]_{\text{meld}}$ with a standard MCMC algorithm would involve computationally infeasible block updates for $\bm{N}^{A}$, $\bm{N}^{G}$, and $\bm{z}$ since $\bm{y}^{A}=f(\bm{N}^{A}, \bm{z})$ and $\bm{y}^{G}=g(\bm{N}^{G})$. We avoided high-dimensional parameter updates by targeting the melded posterior with a multistage MCMC algorithm. We sampled from $[\bm{y}^{A},\cdot]_{\text{ADSM}}$ and $[\bm{y}^{G},\cdot]_{\text{N-mix}}$ using two independent Metropolis-Hastings-within-Gibbs algorithms. We promoted conjugacy of the linear predictor, $\bm{\beta}_{\psi}$, using Pólya-Gamma data augmentation \citep{Polson2013}, which can improve sampling efficiency in ecological binary regression models \citep{Clark2019}. Appendix B includes additional implementation details for the first-stage sampler. 

In the second-stage, density samples from the first-stage were used as the proposals in the STTM (\ref{posteriorSTTM}). For MCMC iteration $k$ in the second-stage, we drew a sub-sample denoted by $\bm{y}^{X,(\star)}$, $X\in\{A, G\}$ from the first-stage samples of submodel $\mathcal{M}$, $\mathcal{M}\in\{\text{ADSM},\text{N-mix}\}$, randomly with replacement, and the Metropolis-Hastings ratio was
\begin{align*}
    & \frac{[\bm{y}^{X,(\star)},\cdot]_{\text{STTM}}[\bm{y}^{X,(\star)},\cdot]_{\mathcal{M}}[\bm{y}^{X, (k-1)},\cdot]_{\mathcal{M}}}{[\bm{y}^{X, (k-1)},\cdot]_{\text{STTM}}[\bm{y}^{X,(k-1)},\cdot]_{\mathcal{M}}[\bm{y}^{X,(\star)},\cdot]_{\mathcal{M}}} = \frac{[\bm{y}^{X,(\star)},\cdot]_{\text{STTM}}}{[\bm{y}^{X, (k-1)},\cdot]_{\text{STTM}}},
\end{align*}
where $\bm{y}^{X, (k-1)}$ is the current value of $\bm{y}^{X}$ in the chain. The refined samples from the second-stage constitute draws from $[\bm{y}^{X},\cdot]_{\text{meld}}$. A heuristic for the multistage MCMC algorithm is that it further refines $[\bm{y}^{X},\cdot]_{\mathcal{M}}$ by selecting samples that conform with the spatio-temporal trends observed in both datasets. To improve mixing, we updated the elements of $\bm{y}^{X}$ one at a time. See Appendix B for a complete description of the second-stage sampler. 

We coded our multistage MCMC algorithm in \texttt{Rcpp} to decrease runtime \citep{Eddelbuettel2011}. The first-stage sampler which targets the posteriors of the ADSM and N-mixture submodel were run in parallel for $100,000$ iterations. We discarded the $10,000$ iterations as burn-in and drew randomly with replacement from the remaining sample for proposals of $\bm{y}^{A}$ and $\bm{y}^{G}$ in the STTM. The second-stage sampler was run for $80,000$ iterations after of burn-in of  $20,000$. Total run times for the first and second stages were $129$ and $111$ minutes, respectively (2.5 Ghz 28-core Intel Xeon W processor). The potential scale reduction factor for all parameters from the first and second stage was less than 1.1 indicating convergence \citep{Gelman1992}.

\section{Results}\label{results}

\begin{figure}
\includegraphics[scale=0.28]{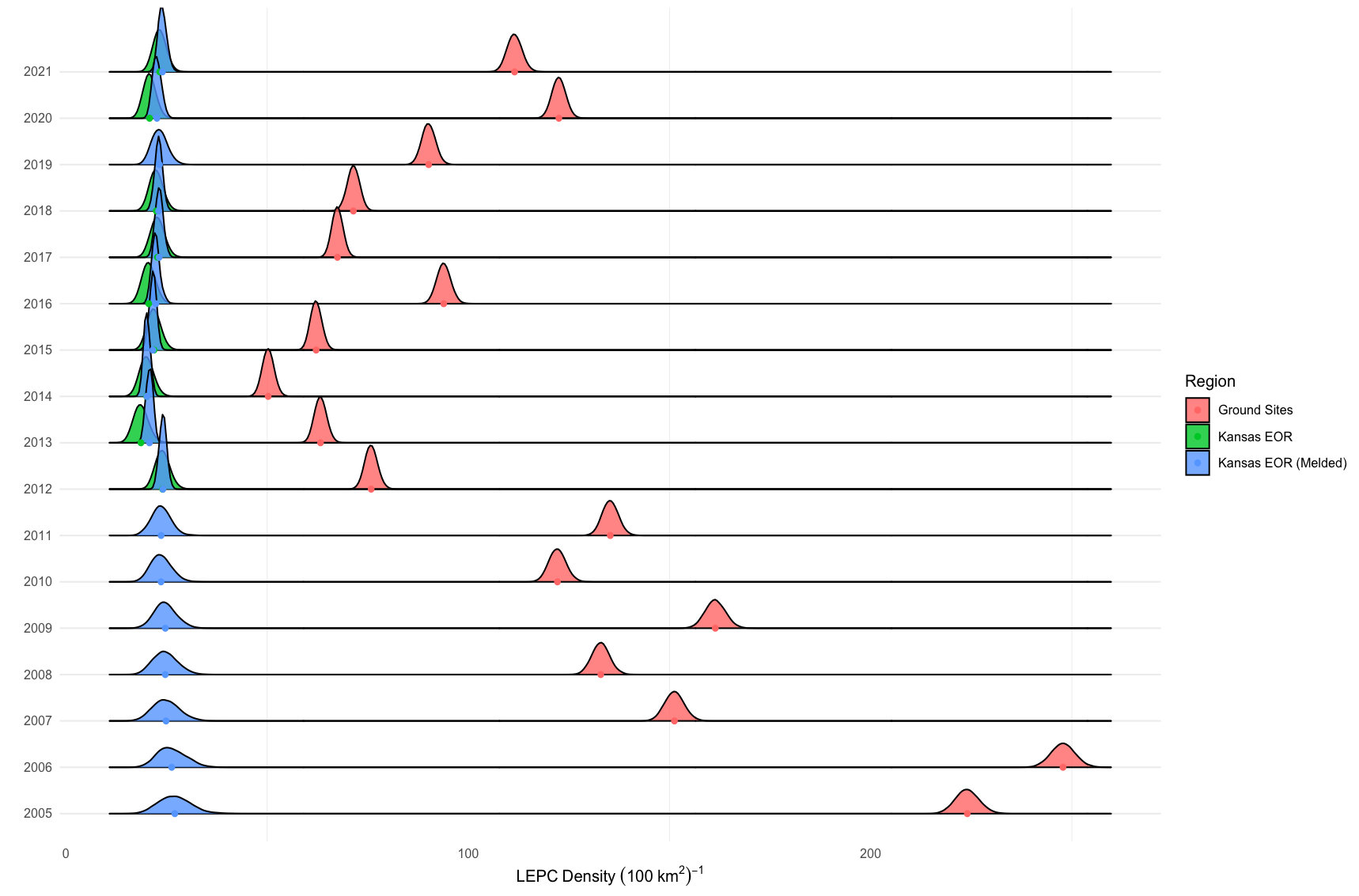} 
\caption{Posterior distributions of annual density for LEPC across Kansas EOR and ground sites from 2005-2021. Red are the posterior annual densities measured across the 21 ground sites estimated from the N-mixture submodel. Green are the posterior annual densities for Kansas EOR estimated from the ADSM. Blue are the refined posteriors for Kansas EOR derived from melding the ADSM and N-mixture submodel densities into the STTM. Posterior means of each distribution are shown as dots.}
\label{fig:denseridges}
\end{figure}

The melded density estimates for the Kansas EOR from the integrated model are similar to the density estimates of the ADSM but have reduced uncertainty and are shifted slightly for some years (Figure \ref{fig:denseridges}). Shifts in the melded posterior tend to mirror trends estimated from the ground surveys. For example, from 2015-2016 there was an estimated decline in LEPC densities according to the aerial survey data, but densities increased across the ground sites. The melded posterior for 2016 incorporates trends from the ground survey and shifts the posterior right. The largest fluctuation in LEPC density was in 2013 following the extreme drought conditions of 2011 and 2012 \citep{Hagen2017}. Both the raw aerial and melded density estimates show a decline, but the fluctuation in the melded estimates is more nuanced. In general, the melded densities estimate have a smoother temporal trajectory compared to the raw aerial estimates. 

\begin{figure}
\includegraphics[scale=0.29]{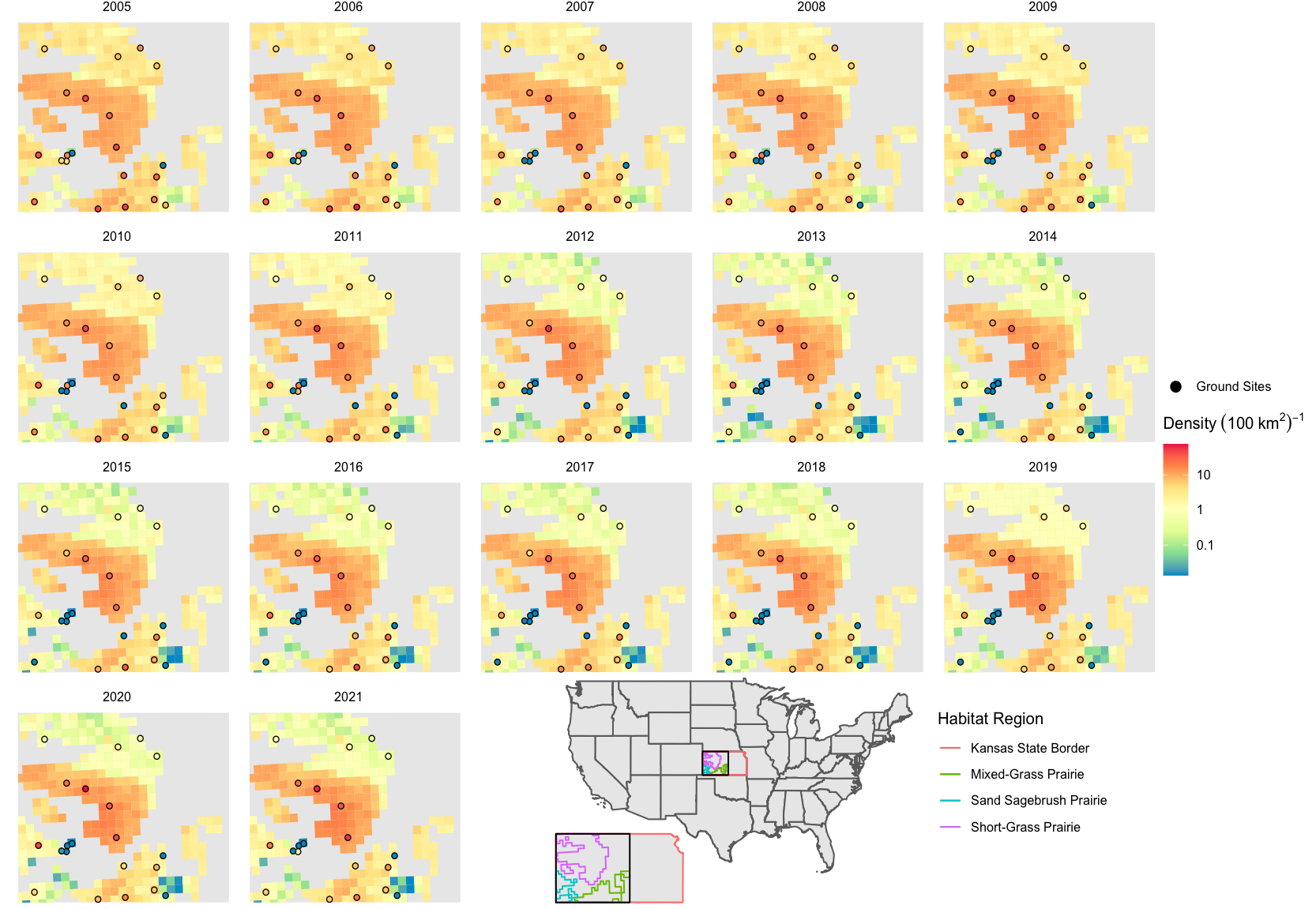} 
\caption{Map of estimated LEPC densities across Kansas EOR and ground sites from 2005-2021. Ecoregions are delineated by outline color in the inset maps. Estimated densities are shown for each survey block along with estimated densities from the 21 ground sites (circles). All densities estimates are from the melded model.
}
\label{fig:densemap}
\end{figure}

The integrated model facilitates inference for LEPC density at unsampled regions via the joint melded distribution so that annual density estimates across Kansas EOR during years which no aerial survey was conducted (2005-2011 and 2019) can still be inferred. The Kansas EOR density estimates from 2005-2011 exhibit greater uncertainty but have long right tails to reflect higher densities observed at the ground survey regions. A map of estimated LEPC across Kansas EOR is given in Figure \ref{fig:densemap}. The southwest region of the SGPR consistently boasted the highest densities followed by the western portion of the MGPR. The SSPR had the lowest densities and show a decreasing pattern over time. Mean estimates were higher in the northern region of the SGPR from 2005-2011 but have large uncertainty because of no aerial or ground surveys during that period (\ref{fig:samplesites}).

\subsection{Simulation Study}\label{simstudy}

We assessed the impacts of Markov melding on predictive performance and inference for a simplified version of our integrated model. Using the STTM in Section \ref{STTM}, we simulated a network of densities at which we generated distance sampling or N-mixture survey data. We rounded the densities simulated from the STTM to the nearest whole number and let that represent the number individuals available for detection at each site. For the aerial sites, we then located simulated individuals uniformly within the survey area. We fit the simulated aerial survey data using a simplified single observer distance sampling model with half-normal detection function. Distances and the parameters of the detection function were specified such that on average, the observer detected half of the individuals in the survey region. See Appendix A for the full model statement. At the ground sites, we set the simulated number of individuals equal to $N_{it}^{G}$ in equation (\ref{N-mixture}) and drew counts for $J=4$ occasions using equation (\ref{N-mixture}) with $p=0.5$. 

We simulated datasets under three different sample size ratios, and also considered datasets simulated with and without preferential sampling. The sample ratios varied from 5 to 20 times more aerial sites than ground sites, with the number of ground sites fixed at $n^{G}=10$. For each sample size ratio, we generated 400 datasets. Each dataset in the STTM consisted of 300 locations. For half of the datasets, we randomly drew a sub-sample of locations for the aerial and ground sites. For the other half, we drew a random sub-sample of aerial sites, but selected the 10 ground sites from the set of 300 that had the highest expected mean density given by $\bm{\zeta}_{0}=\bm{X}_{0}\bm{\gamma}$. The motivation behind our preferential sampling mechanism is that the ground sites in the LEPC case study were opportunistically located based on habitat characteristics known to be associated with higher LEPC density (e.g., large grassland patches and low anthropogenic disturbance). 

\begin{figure}
\includegraphics[scale=0.63]{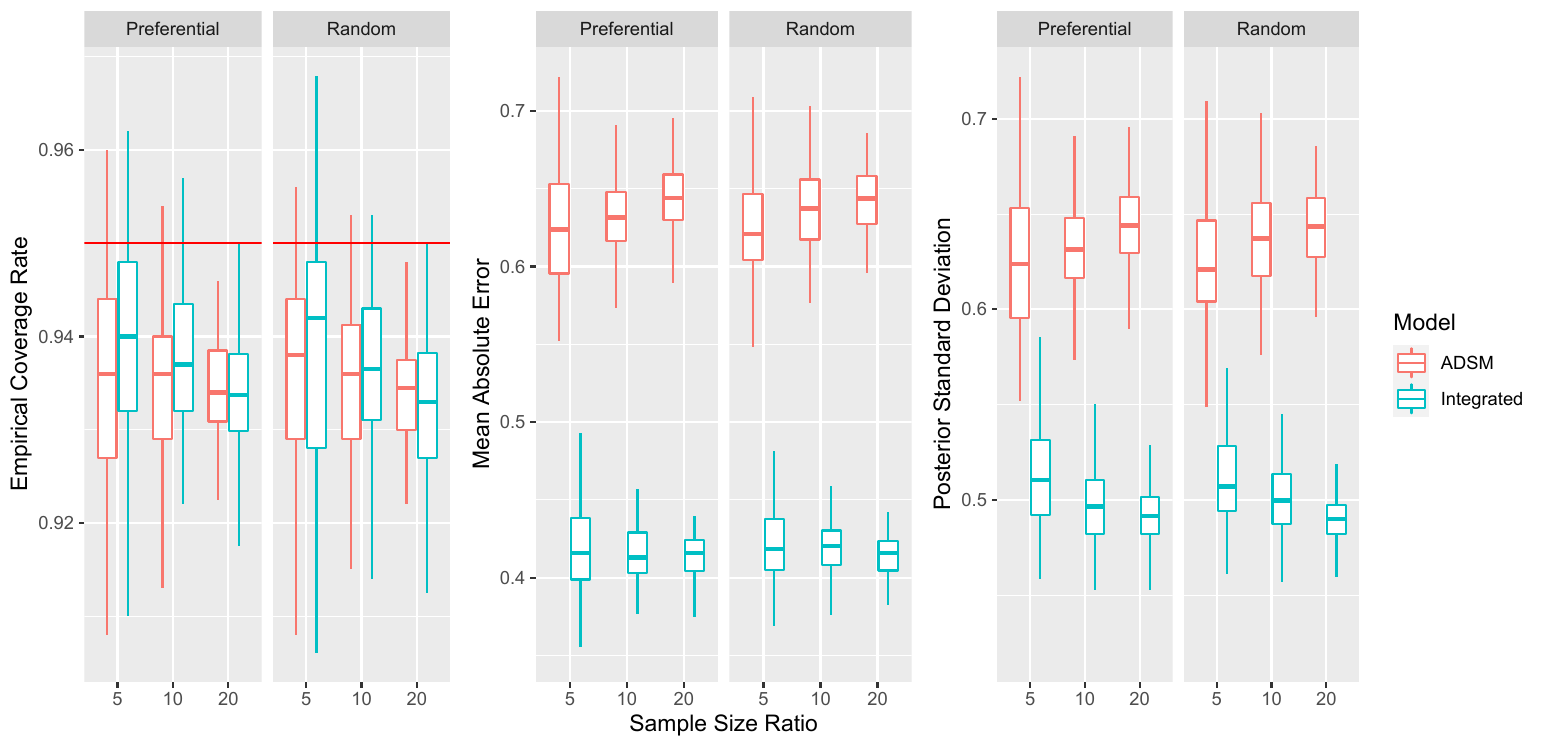} 
\caption{Boxplot of empirical coverage rates, posterior mean absolute errors, and posterior standard deviations for aerial site densities. The red line is the targeted nominal coverage rate of $95$\%. Metrics are given for both the ADSM and full integrated model.}
\label{fig:sim}
\end{figure}

We obtained posterior inference for all $2\times 3\times 200=1200$ datasets using the Markov melding techniques described in Section \ref{modelingfitting} with PoE pooling. The first-stage MCMC algorithm fit the simplified N-mixture and aerial distance sampling submodels in parallel for $10,000$ iterations. The second stage fit the STTM for $20,000$ iterations. For each model fit, we calculated the mean empirical coverage rate, mean absolute error of the posterior mean, and posterior standard deviation of aerial sites densities (Figure \ref{fig:sim}). The results did not differ by sample size ratio or sampling regime. In each case, inference from the integrated model at the aerial sites maintained the same empirical coverage rate of the ADSM but reduced uncertainty and bias. Metrics for the ground sites, not shown, were similar to the aerial sites. The coverage rates for ground sites mimicked those obtained from the N-mixture model in the first-stage but the refined second-stage estimates from the STTM had lower uncertainty and bias. 

\subsection{Sensitivity Analysis}\label{sensitivityanalysis}
 
\begin{figure}
\includegraphics[scale=0.65]{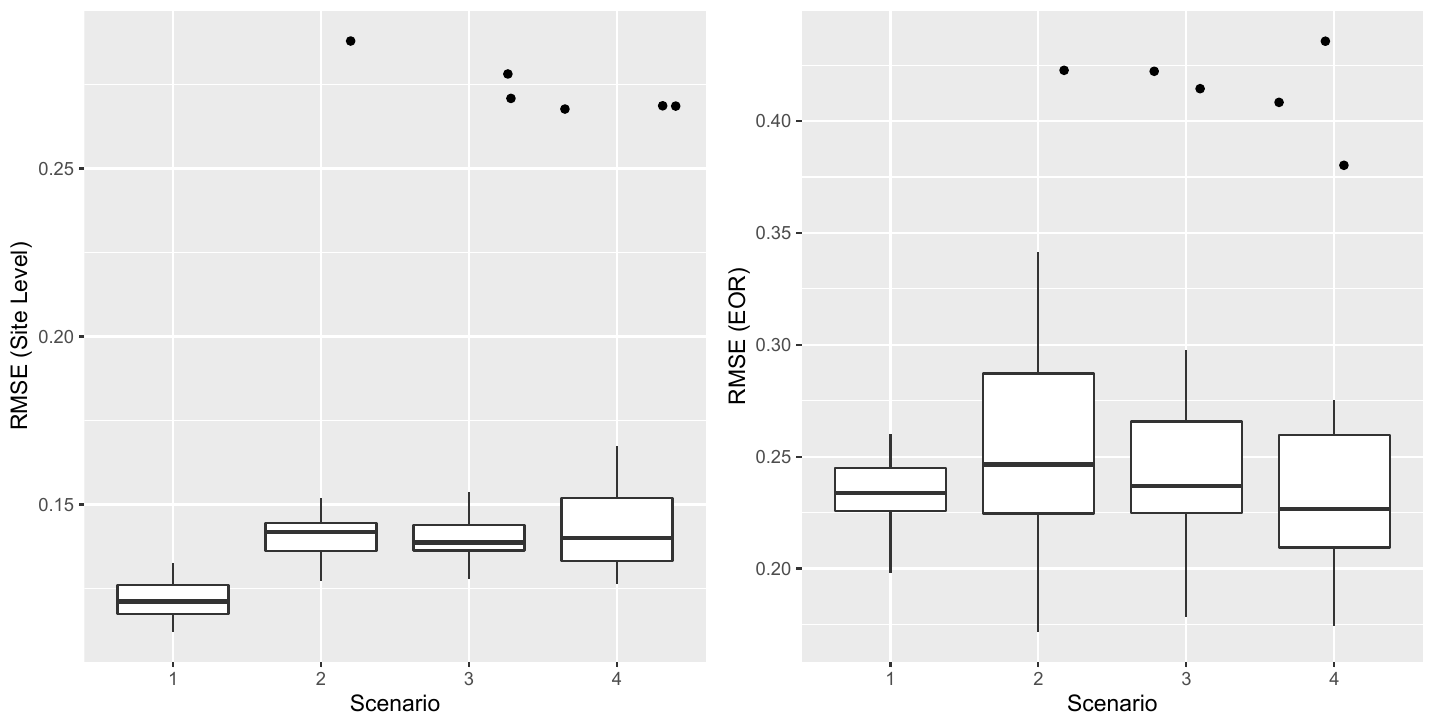} 
\caption{Impact on predictive performance for differing scenarios of missing aerial survey data. In Scenarios 1-4, aerial survey data is available every year, twice every three years, once every two years, and once every three years. The left panel is the RMSE of site-level densities. The right panel is the RMSE of annual abundance predictions divided by the population size.}
\label{fig:sensitivity}
\end{figure}

We performed a sensitivity analysis to assess the inferential cost of conducting aerial surveys less frequently. We considered four different scenarios of missing aerial survey data, but assumed that ground data was available for all ground sites across the 17 years. For each scenario, we simulated $35$ datasets from the STTM using the design and covariance matrices from the LEPC case study. All parameters in the STTM were set to the posterior mean calculated from the fitting the integrated model to the LEPC data. We simulated data for the aerial and ground surveys and fit the integrated model using the same submodels and approach as described in Section \ref{simstudy}. Figure \ref{fig:sensitivity} provides the root mean squared error (RMSE) for site level densities and annual abundances for each scenario. 

All scenarios with missing aerial survey data resulted in substantially higher site level RMSEs than Scenario 1 where aerial surveys were conducted every year. Site level RMSEs were similar for all scenarios for years in which an aerial survey was conducted but much higher in years with no aerial survey because of increased uncertainty. For annual abundance estimation across the EOR, predictive performance was similar across the scenarios with the caveat that Scenarios 2-4 occasionally yielded poor predictive performance. As aerial survey effort decreased, the chances of poor predictive performance increased.
 
\section{Discussion}\label{discussion}

We demonstrated a flexible approach for joint inference from multiple surveys. The need to incorporate mixed surveys into a unified statistical analysis is a common challenge in ecology. Integrated distribution models leverage presence only, detection/nondetection, and count data to infer species latent point patterns \citep{Isaac2020, Simmonds2020}. \cite{Liu2016} developed models for inferring animal trajectories from GPS and ``Dead-Reckoning'' tags. Their model is an adaptation of Bayesian melding models that where originally proposed in atmospheric sciences for linking observations from monitoring stations and the outputs of deterministic climate models to a common Gaussian process \citep{Fuentes2005, Mcmillan2010}. 

Markov melding handles the observational process and spatial support of each data source in separate submodels which can accommodate more complex distributional assumptions. Furthermore, Markov melding facilitates joint inference on quantities that are multivariate non-invertible functions of submodels parameters. This quality is especially appealing in ecology where many popular models provide inference on the parameter of interest through derived quantities. Markov melding may also reduce computation time when submodels handling the observational uncertainty of each dataset are fit in parallel. 

The integrated model reduced uncertainty in annual density estimates by refining the initial density posterior distributions from the submodels to concur with spatio-temporal trends observed across both datasets. Through the melded joint distribution, the integrated model also provided inference for density at unsampled regions that account for the contributions of both datasets. Inferring annual density estimates from the ground sites alone would be inaccurate because of preferential sampling \citep{Diggle2010}. The historical density estimates of the integrated model, however, accounted for the uncertainty in both datasets and leveraged trends in temporal dependence and covariate associations learned from the aerial survey data. The historical density estimates, which provide insights about longer scale trends in LEPC density, are important for assessing recovery of the species \citep{Van2013}.   

Predictive performance of our integrated model varied with aerial survey effort. Overall, site level density estimates were more sensitive to reduced survey effort than range-wide abundance estimates. Reduced aerial survey effort may be adequate for monitoring range-wide populations but could struggle to document fluctuation in the LEPC that are spatially heterogeneous. From a conservation perspective, spatially coarse abundance predictions can be problematic as they have the potential to overlook the contribution of vulnerable subpopulations. 

While range-wide predictive performance was similar across all scenarios, Scenarios 2-4 occasionally performed very poorly. LEPC populations follow a boom-or-bust life history strategy \citep{Ross2016}, which results in large inter-annual variation in abundance that makes prediction difficult. In 2013, a bust was observed that reduced the estimated Kansas LEPC population size by $14$\% (Figure \ref{fig:denseridges}). Without aerial survey data, it would have been difficult to quantify the magnitude of the bust. Thus, reduced aerial effort sampling regimes risk misestimating LEPC boom and busts.

We developed a modeling approach for integrating inference from aerial and ground surveys of LEPC in Kansas, but our approach can be generalized to accommodate other surveys. Most immediately, our approach can accommodate the ground surveys from the other states in the LEPC range. Ground surveys are distinct by state, but each survey produces estimates of density in a particular region and our approach can accommodate differences in observational error. Furthermore, we could extend our current model to account for population dynamics by including an additional submodel that characterizes changes in site-level counts due to annual variability in survival, fecundity, and immigration. The extended IPM could produce spatio-temporal predictions that explicitly account for the contributions of recruitment and survival which could help understand the driver of population change and inform conservation practices for the species \citep{Van2013}.

Accounting for observational error is often a necessity when developing models for SCC \citep{Fernandes2019}. By taking a Markov melding approach, we showed how surveys with unique observational uncertainties and scales can be incorporated into a joint response. Furthermore, we facilitated computation by fitting the model in stages which obviated high-dimensional parameter updates and induced conjugacy for several parameters in the submodels. Another computational advantage of Markov melding is that it enabled model specific data augmentation strategies such as PX-DA in the ADSM and tobit regression in the STTM. Our scalable approach for joint Bayesian inference serves as a foundation for developing future integrated models for mixed surveys of wildlife abundance in other studies.   

\section*{Acknowledgements}
We thank a multitude of landowners in Kansas for allowing access on their properties to conduct the ground counts. The ground survey data were collected by the KDWP. A special thanks to Dana Peterson and Elisabeth Teige for reformatting the Kansas ground survey data for our analysis. We acknowledge the assistance of the WEST crew members and pilots.  \\
\textbf{Data availability}: Data are available upon request from KDWP. \\
\textbf{Funding statement}: The aerial and ground surveys which are the subject of this research article have been financed, in part, with federal funds from the Fish and Wildlife Service, a division of the United States Department of Interior, and administered by the Kansas Department of Wildlife and Parks. The contents and opinions, however, do not necessarily reflect the views or policies of the United States Department of Interior or the Kansas Department of Wildlife and Parks. We thank Charles Rewa and the USDA NRCS as well as Liza Rossi and Colorado Parks and Wildlife for support and funding.   \\


\bibliographystyle{biom}
\bibliography{bibliography.bib}


\section*{Support Information}

Web Appendices referenced in Sections \ref{Models}, \ref{modelingfitting}, and \ref{results} are available with this paper at the Biometrics website on Wiley Online Library. Code for fitting the melded model to the lesser prairie-chicken data are available upon request along with the code for the simulation study and sensitivity analysis. 

\end{document}